# Exploring leakage in dielectric films via automated experiment in scanning probe microscopy

Yongtao Liu,[1] Shelby S. Fields,[2] Takanori Mimura,[2] Kyle P. Kelley,[1] Jon F. Ihlefeld,[2, 3] Sergei V. Kalinin[1]

[1] Center for Nanophase Materials Sciences, Oak Ridge National Laboratory, Oak Ridge, TN 37831, USA

[2] Department of Materials Science and Engineering, University of Virginia, Charlottesville, VA 22904, USA

[3] Charles L Brown Department of Electrical and Computer Engineering, University of Virginia, Charlottesville, VA 22904, USA

Abstract:

Electronic conduction pathways in dielectric thin films are explored using automated experiments in scanning probe microscopy (SPM). Here, we use large field of view scanning to identify the position of localized conductive spots and develop a SPM workflow to probe their dynamic behavior at higher spatial resolution as a function of time, voltage, and scanning process in an automated fashion. Using this approach, we observe the variable behaviors of the conductive spots in a 20 nm thick ferroelectric $Hf_{0.54}Zr_{0.48}O_2$ film, where conductive spots disappear and reappear during continuous scanning. There are also new conductive spots that appear during scanning. The automated workflow is universal and can be integrated into a wide range of microscopy techniques, including SPM, electron microscopy, optical microscopy, and chemical imaging.

Oxide films have found multiple applications in the semiconductor industry. Passive dielectric films are broadly used as a gate dielectric or form sacrificial layers during semiconductor processing. Functional oxides such as ferroelectrics and antiferroelectrics are used for tunable microwave elements, information, or energy storage.[1-3] These applications resulted in a wave of interest towards ferroelectric thin films from the 1980s to the early 2000s. However, traditional perovskite-based ferroelectrics are difficult to integrate into semiconductor processing technology due to thermodynamic instability of the ferroelectric-semiconductor interface. The recent discovery of ferroelectricity in hafnium oxide-based materials has sparked a new wave of research in this area due to its compatibility with semiconductor nanofabrication processes and presence of robust ferroelectricity even in very thin films.[4-6]

However, thin dielectric and ferroelectric films are susceptible to formation of conductive pathways.[7-10] These conductive pathways can be detrimental to the device performance, leading to leakage issues, which necessitate mitigation. Alternatively, these conductive pathways can also be used for designing resistive switching devices. Nonetheless, a complete understanding of the formation and behavior of the conductive pathways in these thin film ferroelectrics will be useful for future development of this class of materials.

Scanning probe microscopy (SPM) is an ideal tool to study the local conductivity of thin film materials.[11] Lau and co-workers have investigated the conductance switching of metal/molecule/metal structures by measuring the local junction conductance using SPM,[12, 13] where they demonstrated that the device switching arises from the appearance and disappearance of nanoscale conducting spots. Williams and coworkers investigated the local conductive channels in other memristor devices and their effects on device performance.[14] Szot *et al.* demonstrated that the switching behavior single-crystalline $SrTiO_3$ originates from local modulations of defects using conductive SPM;[15] these local defects can act as bistable nanowires and hold technological promise for terabit memory devices. Very often these conductive spots, channels, and defects are spatially separated, meaning that they will be essentially localized in large field of view (FoV) microscopy images. Detailed study of such conductive spots requires manual identification and zoomed-in studies.

Here, we developed an automated workflow for scanning probe microscopy (namely, feature-exploration workflow) that enables exploration of specific features in microscopy images, extending previous work on controlled tip trajectories[16] and automated experiment in PFM.[17-22] In the feature-exploration workflow, the features of interest in microscopy images are automatically discovered and located by a machine learning algorithm. Then, the zoomed-in process is performed in an automatic manner according to the parameters of each feature, *e.g.*, the feature size and shape. In the case of studying conductive spots in a thin film, this feature-exploration workflow is integrated with conductive atomic force microscopy (cAFM), abbreviated as FE-cAFM, to investigate local conductivity. Noteworthily, this feature-exploration workflow is compatible with other microscopies.

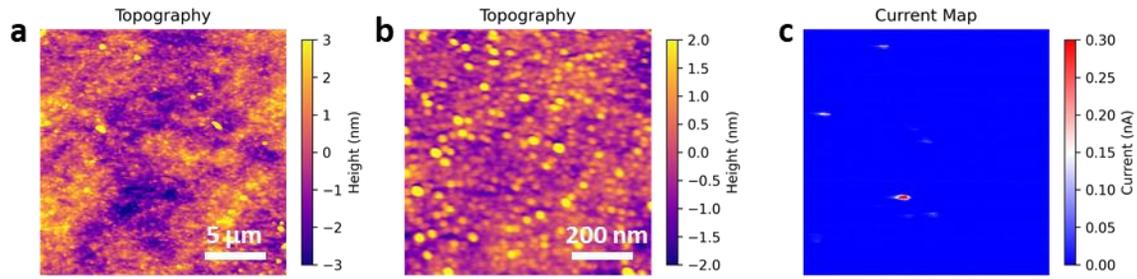

**Figure 1.** cAFM results of the HZO thin film. **(a)**, a large FoV topography image; **(b)**, a small FoV topography image; **(c)**, a cAFM current image corresponding to the small FoV topography in (b).

Here, we integrated this feature-exploration workflow with cAFM to investigate the local conductivity of a 20 nm thick crystalline $Hf_{0.54}Zr_{0.48}O_2$ (HZO) thin film prepared atop a TaN electrode on silicon.[23, 24] The synthesis details of the HZO film is outlined in the Methods section. Shown in Figure 1 are topography and cAFM current images of this HZO thin film. The large FoV topography image in Figure 1a shows the overall morphology, while the small FoV topography image in Figure 1b shows the grain structures of this HZO thin film. The cAFM current image in Figure 1c displays several high conducting spots (red spots). These high-conductivity spots are small (around several tens nanometers) and randomly distributed. The traditional approach to investigate these conductive spots is based on a manual zoomed-in process, which necessitates human intervention to determine the zoom-in scan parameters, *e.g.* zoomed-in location, scan size, etc.

To explore the structure and functionalities of the conductive hot-spots in an automated fashion, we developed a feature-exploration workflow enabling the whole measurement process (from the large FoV scan to small FoV zoom-in scans) to be driven by a computer. Figure 2a shows the schematic of the feature-exploration workflow. The workflow starts with a large FoV AFM measurement, where this FoV can be customized by users based on specific studies. Then, the feature-exploration (the feature is a conductive spot in this work) process is realized via a process of object-segmentation from the background by a threshold filter. The threshold algorithm can be selected from the Scikit-image.[25] According to the location and size of the features, the workflow will make decisions for zoomed-in measurements, *i.e.* generate zoom-in parameters such as location $[x_i, y_i]$, scan size $[s_i]$, scan rate $[r_i]$, etc. The workflow will output and save all results after zoomed-in measurements. Figure 2b is a scheme showing how we implement this workflow in our Cypher SPM, where a field-programmable gate array (FPGA) is connected to the SPM software to realize the real-time control and data acquisition by this feature-exploration workflow.

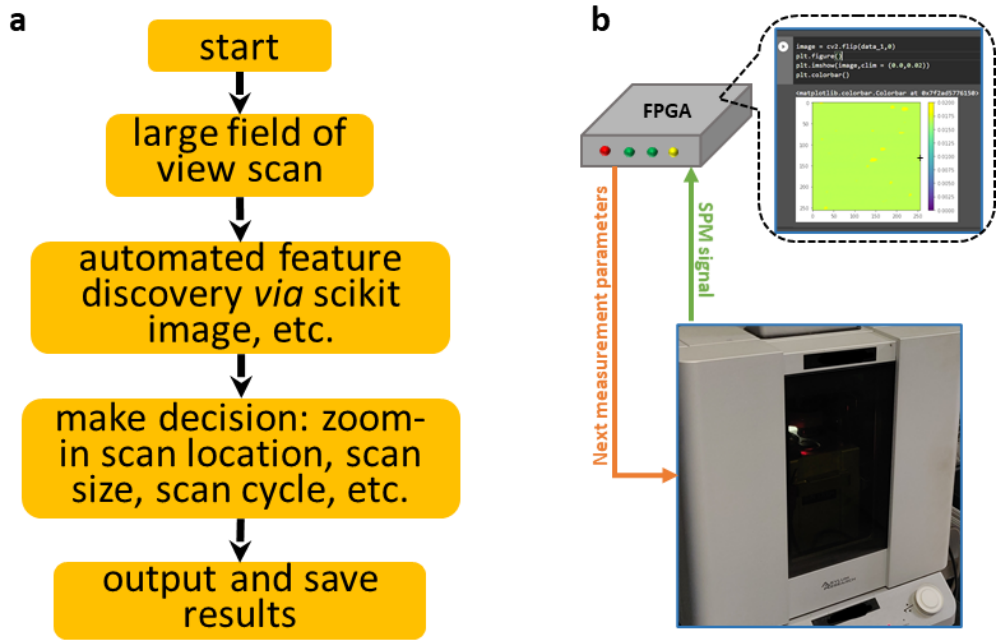

**Figure 2.** Schematic of the automated feature-exploration workflow. **(a)** A scheme illustrating the detailed workflow; **(b)** a scheme showing the intervention between computer and microscopy.

Using FE-cAFM, we investigated the conductive spots in a HZO thin film by applying 6.7 DC voltage to the probe and grounding the bottom electrode, the FE-cAFM results are shown in Figure 3. Figure 3a is a large FoV current image showing four conductive spots. In the segmented image in Figure 3b, the locations of these conductive spots are determined by the feature-exploration workflow, where the marked boxes are the zoom-in measurement regions. Figure 3c is the zoomed-in cAFM results, which show the evolution of these conductive spots as a function of scan cycle in the continuous scan process. Interestingly, these four spots show different behaviors. Both the first and the second conductive spots appear as two separated spots after zoom-in and the shape and size of the conductive spots change during continuous scanning. The second conductive spot disappears in the 4$^{th}$ and 5$^{th}$ zoomed-in scan results (Figure 3c-(2)). A similar phenomenon can also be observed in the zoom-in results of the first conductive spot (Figure 3c-(1)), where the conductive spot is very weak in the 4$^{th}$ zoomed-in image; while differently, the conductive spot becomes stronger again in the 5$^{th}$ zoomed-in image. The third conductive spot (Figure 3c-(3)) is very weak in zoomed-in images, consistent with the behavior in the large FoV image (Figure 3a). Interestingly, the fourth conductive spot completely disappears in the zoomed-in images (Figure 3c-(4)). This FE-cAFM result reveals the different behaviors of the conductive hotspots in this HZO thin film.

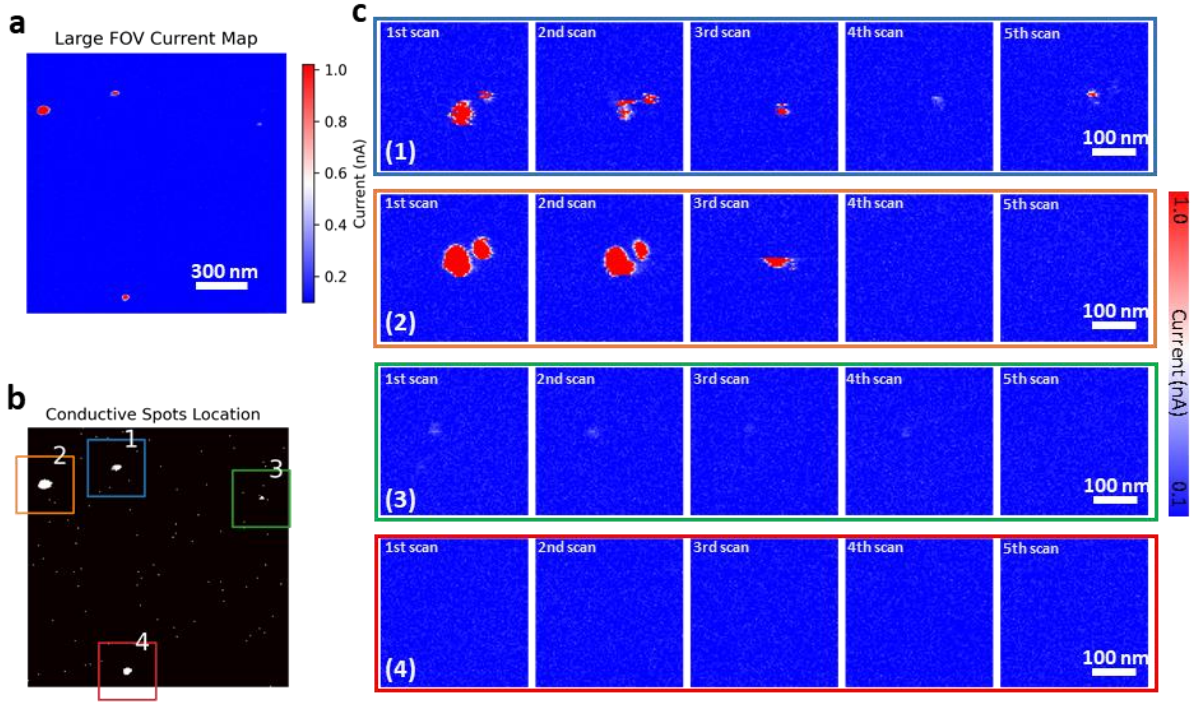

**Figure 3.** FE-cAFM results of the HZO thin film by applying $V_{DC} = 6.7$ V. **(a)** Large FoV current map. **(b)** Conductive spot locations determined by FE workflow. **(c)** Zoomed-in scan results showing the behavior of each conductive spots as a function of scan cycle.

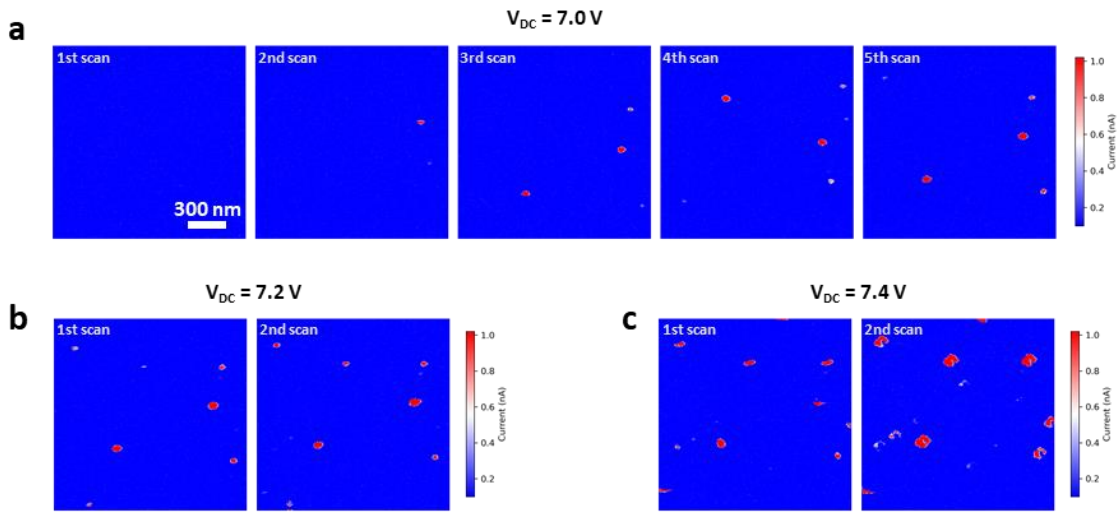

**Figure 4.** The evolution of conductive spots as a function of applied DC voltage. **(a)** cAFM current images showing the evolution of conductive spots under continuous scan with a constant $V_{DC} = 7$ V. **(b)** The evolution of conductive spots under $V_{DC} = 7.2$ V. **(c)** The evolution of the conductive spots under $V_{DC} = 7.4$ V.

To extend our understanding, we also studied the evolution of these conductive spots as a function of applied DC voltage. The results are shown in Figure 4. Shown in Figure 4a are cAFM current images under continuous scanning with a constant $V_{dc}$ = 7.0 V, where we observed the conductive spots gradually appear and/or disappear during scanning. Figure 4b shows the i*n-situ* cAFM current map with an increase of $V_{dc}$ to 7.2 V. Here, we observe more conductive spots compared to the current map in Figure 4a and most spots are larger. In addition, the 2$^{nd}$ scan in Figure 4b also shows larger and higher intensity conductive spots compared to the 1$^{st}$ scan. When we further increased the $V_{dc}$ to 7.4 V (1$^{st}$ scan in Figure 4c), we observed more and enlarged conductive spots again. However, in the 2$^{nd}$ scan under 7.4 V in Figure 4c, we observed a significant change in the spot shape, which motivated us to study the shape of these conductive spots.

Therefore, we performed FE-cAFM measurement based on the cAFM image in the 2$^{nd}$ scan under $V_{dc}$ = 7.4 V (Figure 4c). Corresponding FE-cAFM results are shown in Figure 5. Shown in Figure 5a is a copy of the 2$^{nd}$ scan of Figure 4c, which is used as the large FoV current image for FE-cAFM—this suggests that we can carry out our feature-exploration workflow anytime during SPM measurement, as long as we observed interesting features to explore; in this case, we do not need to start from the beginning of the workflow, we can just use the existing SPM image as the large FoV image for the feature-exploration workflow. Figure 5b shows the locations of the conductive spots determined by the feature-exploration workflow. Notably, we can customize the parameters in the feature-exploration workflow. For example, we changed the parameter of spot size in the feature-exploration workflow, so only relatively large spots are located, as shown in Figure 5b. Shown in Figure 5c are the zoomed-in current images shown the shape of these conductive spots. Interestingly, the "spots" in Figure 5c show a ring shape. Careful inspection of the cAFM measurement suggests that this shape change (from spot to ring) is actually due to SPM tip damage—if the tip edge erodes and the conductive channels in the film are small, then the conductive channel effectively probes the tip shape rather than *vice-versa*.

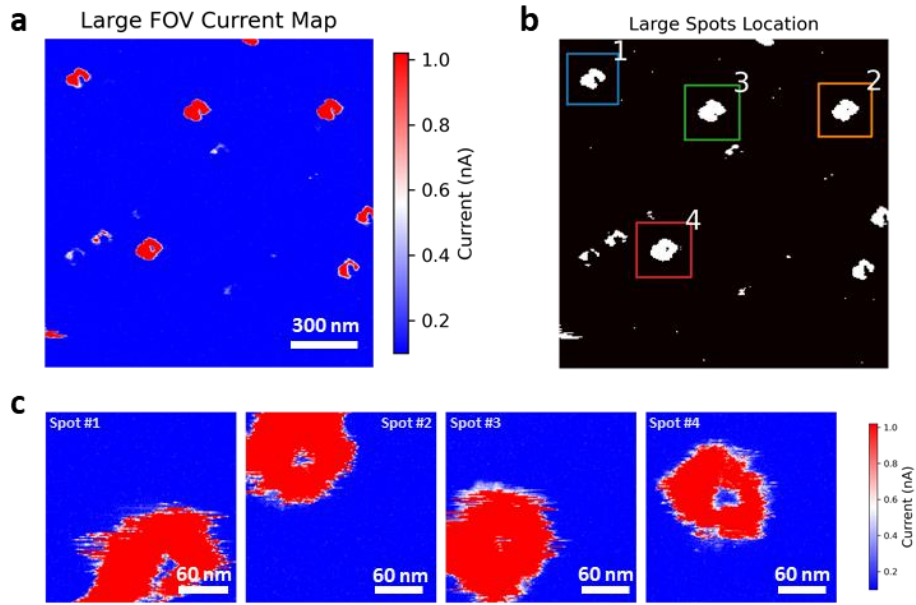

**Figure 5.** FE-cAFM results for exploring the shape change of conductive spots. **(a)** A copy of the cAFM image shown in Figure 4c-1st scan, this image will be used as the large FoV image for the feature-exploration workflow. **(b)** Spot locations determined by the feature-exploration workflow. Here we customize the workflow parameter that only allow it to locate the relatively large spots. **(c)** Zoomed-in scan images showing the shapes of the conductive spots.

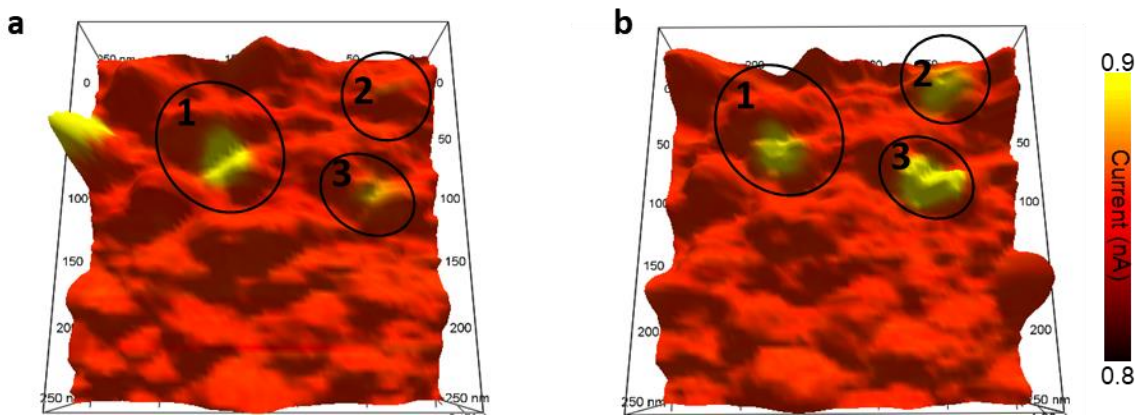

**Figure 6.** Evolution of the conductive spots along with topography; the geography shows topography and the color shows current. **(a),** original topography and current; **(b),** topography and current after continuous scans. **(a)** and **(b)** show a distinct conductive spots change, but there is no corresponding topographic difference. Note that there is an image drift during AFM measurement, so the correspondence between **(a)** and **(b)** is labeled.

We also examined the topography of cAFM measurements, and found no topographic changes corresponding to the evolution of the conductive spots, as shown in Figure 6. In our cAFM measurement, generally we can see more conductive spots under higher $V_{dc}$ (as shown in Figure 4). Therefore, we speculate that these conductive spots are related to defects in the HZO film. We believe that there may exist an asperity in the film that leads to local dielectric breakdown under the high applied fields and the formation of the conductive spots. Possible features include grain boundaries, or high concentrations of oxygen vacancies.

Given that our observation is not identical to several previous works—it was found that the conductivity of $HfO_2$ based films is related to the oxide diffusion, trap density, and dielectric thinning,[26-30] while we do not see compositional variation and thickness variation corresponding to the conductive spots—we suggest that there may exist several different formation mechanisms of the conductive spots. Note that $ZrO_2$ and $HfO_2$ are known fast oxygen ion conductors, if the conductive spots form due to an accumulation of oxygen vacancies that lend themselves to trap assisted tunneling, then the distribution of oxygen vacancies may change after several scans and alter the conduction path. This is also in line with our observation on the evolution of the conductive spots (Figure 3), these behaviors also suggest the difference in the conductive spots. On one hand, the disappearance of the conductive spots suggests a self-healing process of the material in an electric field, which is potentially beneficial to the device; on the other hand, this also indicate the material is changing during operation, implying potential issues of device operation stability. Above all, we believe that more efforts are needed in the future to systematically understand the effect and the mechanism of the conductive spots in HZO films as these materials are poised to make an impact in future computing applications.

In summary, we developed a machine learning based feature-exploration workflow for automated microscopy measurements. This workflow enables the exploration of specific features (*e.g.,* conductive channel, defect, surface contamination, etc.) observable in microscopy images. It locates these features based on a large FoV microscopy image then drives it to perform zoomed-in measurements in an automated manner. By integrating this feature-exploration workflow with conductive atomic force microscopy (FE-cAFM), we investigated the conductive spots in a HZO thin film. The FE-cAFM allowed us to explore the evolution of each conductive spot observed in a large FoV cAFM current image, revealing that the conductive spots appear, disappear, and/or reappear as a function of scan cycles and applied DC voltages. Owing to the capability of the feature-exploration workflow, we discovered variable behaviors of these conductive spots, which may imply their potentially different formation mechanisms.

Noteworthily, we can carry out this feature-exploration workflow anytime during microscopy measurement (that said, not necessarily from the very beginning of the measurement). As long as we observe objects of interest in the microscopy image during measurement, we can immediately carry out the feature-exploration workflow to replace human-guided trivial zoomed-in measurements. In addition, the parameters in this feature-exploration workflow can be customized by users, in doing so, we are able to focus on the specific features of our interest, *e.g.* only focusing on features with specific sizes. This feature-exploration workflow is universal, which can be implemented in any microscopies, including SPM, electron microscopy, optical microscopy, and chemical imaging.

**Methods**

*Feature-exploration workflow*

The detailed methodologies of the feature-exploration workflow are established in Jupyter notebooks and are available from the authors.

*HZO sample preparation*

The 20 nm thick HZO layer was prepared via atomic layer deposition (ALD) atop a 100 nm thick TaN layer that was prepared via DC magnetron sputtering. Full growth details are provided elsewhere.[24] Briefly, the TaN bottom electrode was sputtered from a TaN target onto a p-type (100)-oriented silicon wafer with a native oxide layer. HZO was prepared using plasma-enhanced ALD within an Oxford FlexAL II instrument using TEMA-Hf and TEMA-Zr cation precursors and an oxygen plasma. A supercycle with a Hf:Zr ratio of 6:4 was used to obtain a near $Hf_{0.5}Zr_{0.5}O_2$ composition. Growth was conducted at 260 °C. Following ALD growth, a 20 nm thick TaN layer was deposited using the same DC sputtering process as used for the bottom electrode. The entire film stack was then rapid thermal processed at 600 °C for 30 seconds in dynamic $N_2$ to induce crystallization. Following processing, the top electrode layer was removed via an SC-1 chemical treatment (1:1:5 $NH_4OH:H_2O_2:H_2O$, 58 °C, 45 minutes).

*cAFM measurements*

The cAFM was performed on an Oxford Instrument Asylum Research Cypher microscope using Budget Sensor Multi75E-G Cr/Pt coated AFM probes (~3 N/m).

**Conflict of Interest**

The authors declare no conflict of interest.

**Authors Contribution**

S.V.K. and Y.L. conceived the project. Y.L. designed the workflow. Y.L. and K.P.K implemented it in the Cypher SPM based on the previous efforts from K.P.K.. S.S.F., T.M., and J.F.I. synthesized the HZO sample. All authors contributed to discussions and the final manuscript.

**Acknowledgements**

This effort was supported by the center for 3D Ferroelectric Microelectronics (3DFeM), an Energy Frontier Research Center funded by the U.S. Department of Energy (DOE), Office of Science, Basic Energy Sciences under Award Number DE-SC0021118, and the Oak Ridge National Laboratory's Center for Nanophase Materials Sciences (CNMS), a U.S. Department of Energy, Office of Science User Facility. The authors acknowledge Stephen Jesse and Nina Balke for fruitful discussion.